\documentclass[sigconf]{acmart}
\AtBeginDocument{%
  }

\copyrightyear{2026}
\acmYear{2026}
\setcopyright{cc}
\setcctype{by}
\acmConference[WWW '26]{Proceedings of the ACM Web Conference 2026}{April 13--17, 2026}{Dubai, United Arab Emirates}
\acmBooktitle{Proceedings of the ACM Web Conference 2026 (WWW '26), April 13--17, 2026, Dubai, United Arab Emirates}
\acmPrice{}
\acmDOI{10.1145/3774904.3792862}
\acmISBN{979-8-4007-2307-0/2026/04}

\usepackage{tabularx} 
\usepackage{booktabs} 
\usepackage{array}
\usepackage{balance} 

\begin{document}

\title{Generative Retrieval for E‑commerce: Jointly Learning Embedding and Codebook with Same Product Cluster}

\author{Songtao Fang}
\orcid{0009-0001-7713-0415}
\authornote{Corresponding author.}
\affiliation{%
  \institution{Alibaba Group}
  \city{HangZhou}
  \country{CHINA}
}
\email{fangsongtao.fst@taobao.com}

\author{Zihao Xu}
\orcid{0009-0002-2973-8183}
\affiliation{%
  \institution{Alibaba Group}
  \city{HangZhou}
  \country{CHINA}
}
\email{xiyu.xzh@taobao.com}

\author{Shaowei Wei}
\orcid{0009-0004-9473-0895}
\affiliation{%
  \institution{Alibaba Group}
  \city{HangZhou}
  \country{CHINA}
}
\email{leqian.wsw@taobao.com}
\author{Jin Zhang}
\orcid{0009-0006-4073-7869}
\affiliation{%
  \institution{Alibaba Group}
  \city{HangZhou}
  \country{CHINA}
}
\email{zj146372@taobao.com}
\author{Zhuojun Wang}
\orcid{0009-0008-8685-4467}
\affiliation{%
  \institution{Alibaba Group}
  \city{HangZhou}
  \country{CHINA}
}
\email{jerry.wangzj@taobao.com}

\renewcommand{\shortauthors}{Songtao Fang, Zihao Xu, Shaowei Wei, Jin Zhang, \& Zhuojun Wang}

\begin{abstract}
With the development of large language models (LLMs), generative retrieval is becoming increasingly important in e-commerce scenarios. Current mainstream approaches typically use a two-stage training strategy: first train a product embedding model, and then learn a codebook that maps embeddings to product IDs. This cascaded approach suffers from two major issues: (1) error accumulation—if the embedding model in the first stage produces biased representations, the codebook in the second stage cannot correct these errors, degrading final retrieval performance; and (2) codebook learning relies solely on product embeddings and lacks modeling of query-to-product and product-to-product interactions. As a result, products belonging to the same cluster may be assigned inconsistent IDs by the codebook, further hurting retrieval accuracy. To address these problems, we propose a novel method that jointly trains the embedding model and the codebook, and incorporates same product cluster information as an additional supervision signal. Experimental results demonstrate that our method significantly improves e-commerce retrieval performance while simultaneously enhancing both embedding and codebook learning.
\end{abstract}




\maketitle

\section{Introduction}
Generative retrieval (GR) has emerged as a promising approach in information retrieval. Unlike sparse or dense methods—which rely on shallow semantic matching and often fail to capture implicit intent\cite{shao2025reasonir,su2024bright} or handle colloquial queries (e.g., “gifts for a 10-year-old who loves science”). GR leverages large generative models to better interpret user intent and produce more relevant results.

GR can directly produce document identifiers (DocIDs) from user queries using the capabilities of pre-trained models. DPI\cite{tay2022transformer} introduced transformer-based autoregressive models that preprocess documents into hierarchical identifiers using k-means clustering. Other approaches, such as MERGE\cite{zhang2025multi}, LETTER\cite{wang2024learnable}, Tiger\cite{rajput2023recommender}, and UniSearch\cite{chen2025unisearch}, learn DocIDs for retrieval based on RQ-VAE\cite{rajput2023recommender} or VQ-VAE\cite{van2017neural}.
However, mainstream approaches employ a two-stage pipeline: first learning product embeddings, then deriving a codebook to map these embeddings to DocIDs. This introduces two fundamental limitations. First, error accumulation: inaccuracies in the learned embeddings are propagated and amplified during codebook training. Second, insufficient interaction modeling: the codebook relies solely on static embeddings and thus fails to capture query–product and inter-product interactions, often resulting in inconsistent product IDs assignments even among semantically equivalent products within the same product cluster (i.e., variations of the same product from different suppliers or in different sizes).

To address these issues, we propose a novel joint training framework that simultaneously optimizes both the product embedding and the codebook. Our framework uses product cluster assignments as auxiliary supervision, which explicitly enhances the semantic consistency among products within the same cluster. By jointly training both components, our approach eliminates the error propagation inherent in conventional two-stage pipelines and optimizes them under a unified objective. 
The main contributions of this paper are as follows:
\begin{enumerate}
    \item We propose a joint training framework to optimize product embedding and the codebook, addressing error accumulation in traditional two-stage methods.
    \item By leveraging product cluster information as additional supervision, we enhance both the semantic consistency of IDs and the quality of embeddings for identical products.
    \item Extensive experiments on multiple e-commerce retrieval tasks show that our method outperforms existing approaches.
\end{enumerate}
\begin{figure*}[t]
  \centering
  \includegraphics[width=\textwidth, height=0.4\textheight, keepaspectratio]{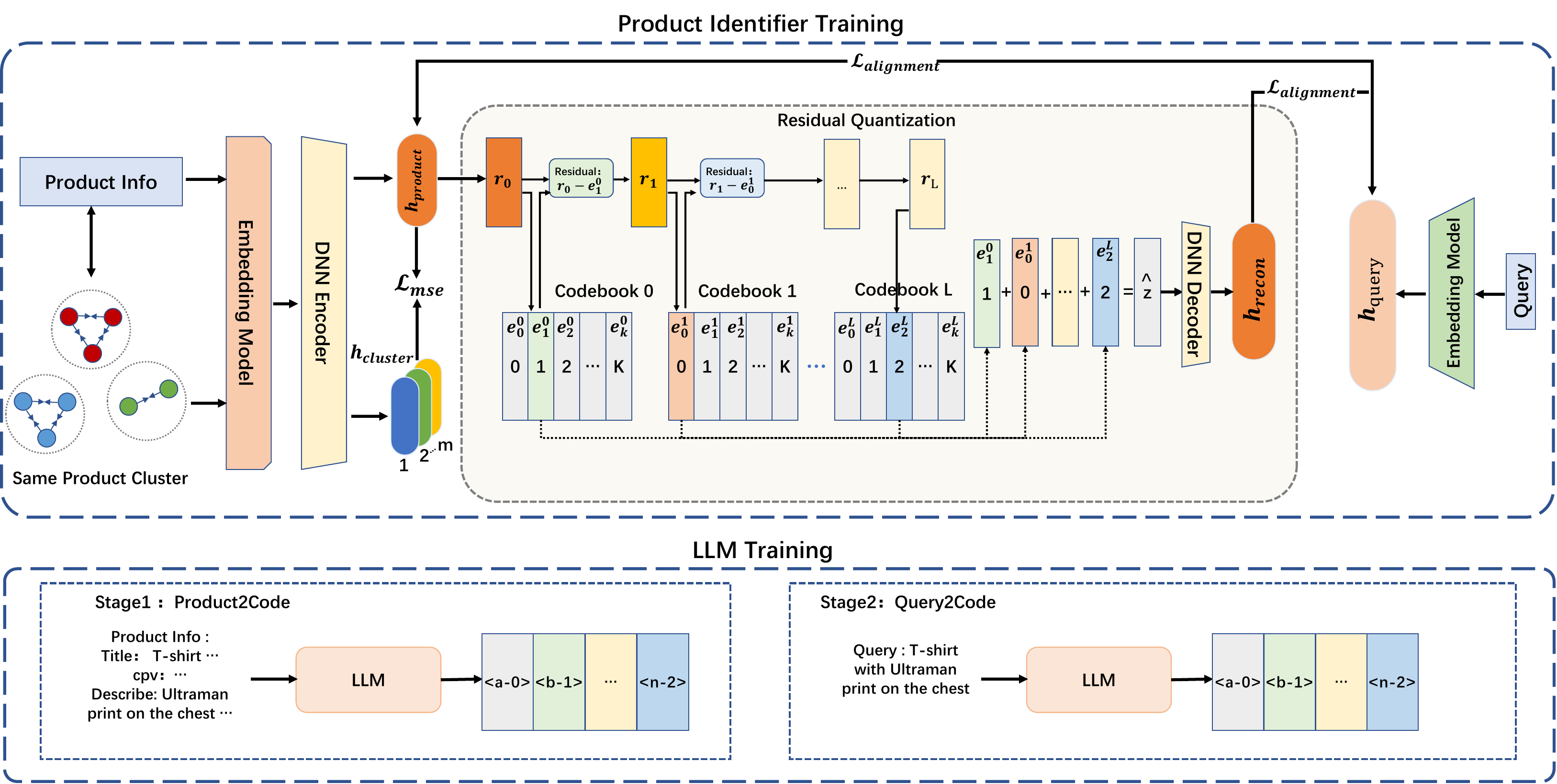}
  \caption{Architecture of our model, consisting of Product Identifier Training and LLM Training.}
  \label{fig:1}
\end{figure*}
\section{Related Work}
\noindent \textbf{Sparse and Dense Retrieval.}
Sparse retrieval methods like BM25\cite{robertson2009probabilistic} use inverted indexes and TF‑IDF–style term matching for efficient lookup, but they struggle with semantic variation and term mismatch. Dense retrieval approaches such as DPR\cite{karpukhin2020dense} and ColBERT\cite{khattab2020colbert} leverage neural embeddings and pretrained models like BERT\cite{devlin2019bert} to capture semantic relationships and improve accuracy. However, both rely on fixed features or learned representations that can fail to capture complex query intent. Large language models, by contrast, excel at understanding complex queries—an advantage that neither sparse nor dense retrieval consistently provides. 

\noindent \textbf{Generative Retrieval.}
Generative retrieval has emerged as a promising paradigm in information retrieval. DSI\cite{tay2022transformer} transforms documents into DocIDs and utilizes transformers for end-to-end retrieval. GenRet\cite{sun2023learning} uses an Encoder-Decoder structure to generate ID tokens step-by-step. MERGE\cite{zhang2025multi} introduces a novel framework that employs multi-level relevance learning within an RQ-VAE to generate document identifiers. GRAM\cite{pang2025generative} introduces a novel approach to e-commerce retrieval, leveraging large language models to generate shared codes for both queries and products. Mainstream methods typically adopt a two-stage pipeline, first learning embeddings and then mapping them to product IDs, which often leads to error accumulation and lacks effective interaction modeling.
\section{Model}

In this section, we present our framework, including the Product Identifier Training and the LLM Training. The product retrieval task can be formalized as the process of retrieving a relevant product \(p\) for a user query \(q\) from a large catalog of products $\mathcal{P} $. Each product $p \in \mathcal{P} $is associated with rich multimodal attributes, such as title, description, and category, which can be represented as a token sequence $t = \left\{ t_1, \ldots, t_{|d|} \right\}$where $|d|$ is the total number of tokens in the product info. To enable efficient end-to-end retrieval, we first adopt a product tokenization strategy that maps each product \(p\) to a discrete token sequence $z = \left\{ z_1, \ldots ,z_l, \ldots, z_L \right\}$ , where each token \(z_l\) is a \(K\)-way categorical variable($z_l \in \left\{ 1, 2, \ldots, K \right\}$, and \(L\) denotes the fixed length of the product identifier. Crucially, each product \(p\) also belongs to a group of the same product \(c(p)\) \(c(p)\)—a group of functionally or visually identical items (e.g., different sellers offering the same SKU)—which provides valuable semantic supervision. Then we train an LLM to learn the mapping from user queries to product IDs.
\subsection{Product Identifier Training}
The Product Identifier Training is used to jointly train embeddings and codebooks.

\noindent \textbf{Query\&Product Encoding.}Given a query and the corresponding product  \(p\), we first extract their semantic features \(h_{product}\) and \(h_{query}\) using an embedding model and a deep neural network(DNN) encoder. Additionally, we randomly sample \(m\) products from the same product cluster \(c(p)\) to which the product belongs, extract their semantic features $h_1, h_2, \ldots, h_m$, and compute their mean to obtain \(h_{cluster}\). Mathematically, this can be expressed as follows:
\begin{equation}
h_{\text{cluster}} = \frac{1}{m} \sum_{i=1}^{m} h_i
\end{equation}
Then we apply the Mean Squared Error (MSE) loss to enforce consistency between the product embedding \(h_{product}\) and the cluster-level representation \(h_{cluster}\):
\begin{equation}
\mathcal{L}_{mse}=MSE(h_{product},h_{cluster})
\end{equation}
\noindent \textbf{Semantic Embedding Residual Quantization.}The latent semantic embedding \(h_{product}\)is quantized into a code sequence using a hierarchy of codebooks consisting of \(L\) levels , where \(L\) represents the length of the identifier. At each quantization level \(l\), there is a codebook $ C_l = \{ e_k^l \}_{k=1}^K$, where $e_k^l \in \mathbb{R}^d$ represents a learnable code embedding and \(K\) denotes the codebook size. At each code level, we find the most similar code embedding with the semantic residual and assigns the product with the corresponding code index. The residual quantization can be formulated as:
\begin{equation}
\begin{cases}
z_l = \arg\min \| \mathbf{r}_{l-1} - \mathbf{e}_k^l \|^2, \quad \mathbf{e}_k^l \in C_l, \\
\mathbf{r}_l = \mathbf{r}_{l-1} - \mathbf{e}_{z_l}
\end{cases}
\end{equation}
where, \(z_l\) is the code index chosen from the \(l\)-th codebook, \(r_{l-1}\) is the semantic residual from the last level. In the first step, we set \(r_0\)=\(h_{product}\). After the residual quantization, we obtain the quantized identifier $z = [z_1, \ldots ,z_l, \ldots, z_L]$ and the quantized representation $\hat{z} = \sum_{l=0}^{L-1} e_{z_l}$. The quantized representation is fed into a DNN decoder \(D\) ($h_{recon}=D(\hat{z})$) to reconstruct the input \(h_{product}\) via a reconstruction loss:
\begin{equation}
\mathcal{L}_{recon}=||h_{product}-h_{recon}||^2
\end{equation}
The $\mathcal{L}_{RQ-VAE}$ objective combines reconstruction loss $\mathcal{L}_{recon}$ with codebook commitment loss $\mathcal{L}_{rq}$:
\begin{equation}
\begin{cases}
\mathcal{L}_{RQ-VAE}=\mathcal{L}_{recon} + \mathcal{L}_{rq},     where, \\
\mathcal{L}_{rq}=\sum_{l=0}^{L}(||sg[r_l]-e_{z_l}^l||^2 + \mu||r_l-sg[e_{z_l}^l]||^2)
\end{cases}
\end{equation}
where \(sg[\cdot]\) is the stop-gradient operation, and \(\mu\) is the coefficient to balance the strength between the optimization of code embeddings and encoder.

\noindent \textbf{Contrastive Learning.} To achieve joint training of the codebook and embedding model, we utilize InfoNCE loss\cite{oord2018representation} on two key components: the representation \(h_{product}\)
generated by the embedding model and the reconstructed embedding representations \(h_{recon}\). Both components share the same InfoNCE learning objective:

\begin{equation}
\mathcal{L}_{\text{InfoNCE}} =
    -\log \frac{\exp\left( s(h_{query}, p^+)/\tau \right)}
    {\sum_{j=1}^{N}  \exp\left( s(h_{query}, p^i)/\tau \right)} 
\end{equation}
where \(\tau\), \(h_{query}\), and \(p\) denote the temperature parameter, query representation and product representation. In the two components, \(p\) represents \(h_{product}\) and \(h_{recon}\), respectively. The positive \(p^+\)is the relevant product to query, and other irrelevant product are negatives. These negatives can be either hard-negatives or in-batch negatives (products of other instances in the same batch). \(s(\cdot)\) is the relevance score of query and product, measured by the cosine similarity between their respective representations.

\noindent \textbf{Overall Loss.} The training loss for our model is summarized as:
\begin{equation}
\mathcal{L}_{\text{model}} = \mathcal{L}_{RQ-VAE} + \alpha\mathcal{L}_{\text{InfoNCE}} + \beta\mathcal{L}_{mse}
\end{equation}
where \(\alpha\) and \(\beta\) are hyper-parameters to control the strength of contrastive learning and same product cluster constraints, respectively.
\subsection{LLM Training}
LLM Training is designed to learn the mapping from user queries to product IDs. Assuming the length of the product identifier is 4, and using the trained embedding and VAE model, each product is assigned an ID sequence such as: [<a-1>,<b-0>,<c-6>,<d-2>]. Each element in this sequence, such as <a-1>, represents a distinct special token added to the vocabulary for training. Since the newly introduced special tokens are unfamiliar to LLMs, we adopt a two-stage progressive learning strategy to better achieve semantic alignment and remodeling. In the first stage, we train the model to map product information (such as titles, attributes, and product descriptions) to product IDs, enabling it to preliminarily grasp the semantic information of the new tokens. In the second stage, we further train the model to map user queries to product IDs. Through this staged learning approach, the model can more effectively master the semantic characteristics of the new tokens, thereby improving retrieval performance. Finally, we train the sequence-to-sequence transformer model with the following loss function:
\begin{equation}
\mathcal{L}_{sft} = -\sum_{t=1}^{T}log P(y_t|y_{<t},input)
\end{equation}
where the input represents product information in Stage 1, and represents user query in Stage 2. \(T\) denotes the length of product ID, \(y_t\) is the \(t\)-th token of the target product ID, \(y_{<t}\) represents the previously generated tokens.
\section{Experiments}
\subsection{Datasets Description}
We use 20 million product data from Alibaba's internal e-commerce platform to train embeddings and codebooks. Similarly, we use these 20 million products to learn product-to-ID mappings for LLM stage1 training. Additionally, we construct a mapping of 40 million queries to product IDs for stage2 training. We also constructed validation and test sets, each containing 10,000 queries, evenly distributed between regular and implicit intent queries to ensure comprehensive model evaluation.
\subsection{Baseline Models}
To investigate the effectiveness of our model, we use three baseline categories: sparse, dense, and generative retrieval models. The detailed introductions are listed as follows: 
\begin{itemize}
    \item BM25\cite{robertson2009probabilistic}: It enhances relevance scoring by refining term weights based on the TF-IDF feature.
    \item DPR\cite{karpukhin2020dense}: This approach enhances retrieval effectiveness by capturing deeper contextual relationships between queries and documents. To further support long text sequences, we utilize the GTE embedding \cite{zhang2024mgte} as encoder.
    \item DSI\cite{tay2022transformer}: It represents documents using hierarchical K-means clustering results.
    \item $\mathrm{Tiger}_{\text{rq-vae}}$\cite{rajput2023recommender}: It quantizes semantic information into code sequence for LLM-based generative recommendation. Here, we focus solely on comparing the RQ-VAE component. 
\end{itemize}

\begin{table}[htbp]
    \centering 
    \caption{Performance comparison of different models} 
    \label{tab:experiment_results}
    \begin{tabular*}{\linewidth}{@{\extracolsep{\fill}}lcccc} 
        \toprule
        Model & Recall@1 & Recall@10 & Recall@100 & ALSP\\
        \midrule
        BM25 & 1.23 & 2.66 &  9.80 & \textemdash \\ 
        DPR  & 3.57 & 7.09 & 24.24 & \textemdash \\
        DSI &  3.79 & 7.99 & 23.91 & 2.89 \\
        $\mathrm{Tiger}_{\text{rq-vae}}$ &  4.33 & 8.84 & 26.38 & 3.92 \\
        ours & \textbf{4.49} & \textbf{9.90} & \textbf{30.71} & \textbf{4.42} \\
        ours-cluster  & 4.39 & 8.91 & 28.47 & 4.01 \\
        \bottomrule
    \end{tabular*}
\end{table}

\subsection{Evaluation Metrics}
We evaluate our model using Recall@K(\%), which measures the
fraction of relevant items retrieved in the top-K
results, reflecting the model’s ability to identify
relevant candidates. We also evaluate the quality of the codebook using the \textbf{A}verage \textbf{L}ength of the \textbf{S}hared ID \textbf{P}refixes (ALSP) among items within the same product cluster. 
\subsection{Experiment Settings}
For product identifier training, we utilize the gte embedding model\cite{zhang2024mgte} as the shared query and product embedding model. In our RQ-VAE implementation, the codebook is configured with 5 layers (\(L\)=5), each containing 32 entries (\(K\)=32), and the embedding size for each entry is set to 768. Additionally, up to 5 products are sampled from the same product cluster corresponding to each product and we use the AdamW
optimizer with a learning rate of 1e-4. We follow\cite{rajput2023recommender} to set \(\mu\) as 0.25, \(\alpha\) is set as 0.5 \(\mu\) and \(\beta\) is set to as 0.5. For LLM training, we employ Qwen2.5-7B\cite{team2024qwen2} to learn the mapping from product and query to product IDs with a learning rate of 5e-5. 

\subsection{Results and Analysis}
The experimental results are shown
in Table~\ref{tab:experiment_results}. ours-cluster denotes our model with the product-cluster constraint removed. Overall, the experimental results indicate that the proposed model outperforms all baselines  on a large-scale real-world dataset. Specifically, compared to sparse (e.g., BM25) and dense retrieval methods (e.g., DPR), our model significantly outperforms them on the dataset. Sparse retrieval relies on keyword matching, while dense retrieval is constrained by the construction of positive and negative samples and cannot leverage general world knowledge. These limitations make both approaches struggle to handle complex or implicit-intent queries effectively.

Compared with generative retrieval methods, our model exhibits superior effectiveness and practicality. DSI and $\mathrm{Tiger}_{\text{rq-vae}}$, which rely on item embeddings for clustering and residual quantization, tend to amplify inherent biases in the embedding space. Furthermore, their non-end-to-end learning paradigm introduces information loss, degrading overall retrieval performance. As evidenced by the ASLP metric, items within the same product cluster exhibit significantly higher common prefixes than those generated by other generative methods. Crucially, when the cluster-based vector constraints are removed, both ASLP and other key metrics exhibit substantial degradation. This demonstrates that the product-cluster constraint provides a powerful, semantically rich supervision signal for learning consistent item ID representations. By aligning item IDs with meaningful semantic groupings, our approach reduces ambiguity in subsequent LLM training, enabling more coherent and hierarchical semantic structures in the ID space. By comparing the performance of ours-cluster with $\mathrm{Tiger}_{\text{rq-vae}}$, we can observe that end-to-end training brings performance improvements, as the interaction between queries and products during training helps mitigate the bias inherent in codebook learning that relies solely on embeddings.

\section{Conclusion}
In this paper, we propose a joint-training framework for generative retrieval that simultaneously optimizes product embedding and codebook, using same product cluster information to enforce ID consistency. Our method eliminates error accumulation in two-stage pipelines and significantly outperforms existing approaches on real-world e-commerce data. Ablations confirm that both joint training and cluster supervision are crucial for enhancing retrieval accuracy and improving product ID consistency within the same cluster.

\bibliographystyle{ACM-Reference-Format}
\bibliography{sample-base}

\appendix

\end{document}